\documentclass[conference]{IEEEtran}
\IEEEoverridecommandlockouts
\usepackage{cite}
\usepackage{amsmath,amssymb,amsfonts}
\usepackage{algorithmic}
\usepackage{graphicx}
\usepackage{textcomp}
\usepackage{xcolor}
\def\BibTeX{{\rm B\kern-.05em{\sc i\kern-.025em b}\kern-.08em
    T\kern-.1667em\lower.7ex\hbox{E}\kern-.125emX}}
\begin{document}

\title{A Novel Zero-Trust Machine Learning Green Architecture for Healthcare IoT Cybersecurity: Review, Analysis, and Implementation\\
}

\author{\IEEEauthorblockN{Zag ElSayed}
\IEEEauthorblockA{\textit{School of Information Technology} \\
\textit{CECH, University of Cincinnati}\\
Ohio, USA \\
}
\and
\IEEEauthorblockN{Nelly Elsayed}
\IEEEauthorblockA{\textit{School of Information Technology} \\
\textit{CECH, University of Cincinnati}\\
Ohio, USA \\
}
\and
\IEEEauthorblockN{Sajjad Bay}
\IEEEauthorblockA{\textit{Architecture and Data Science Dept.} \\
\textit{Johnson Controls}\\
Wisconsin, USA \\
}
}

\maketitle

\begin{abstract}
The integration of Internet of Things (IoT) devices in healthcare applications has revolutionized patient care, monitoring, and data management. The Global IoT in Healthcare Market value is \$252.2 Billion in 2023. However, the rapid involvement of these devices brings information security concerns that pose critical threats to patient privacy and the integrity of healthcare data. This paper introduces a novel machine learning (ML) based architecture explicitly designed to address and mitigate security vulnerabilities in IoT devices within healthcare applications. By leveraging advanced convolution ML architecture, the proposed architecture aims to proactively monitor and detect potential threats, ensuring the confidentiality and integrity of sensitive healthcare information while minimizing the cost and increasing the portability specialized for healthcare and emergency environments. The experimental results underscore the accuracy of up to 93.6\% for predicting various attacks based on the results demonstrate a zero-day detection accuracy simulated using the CICIoT2023 dataset and reduces the cost by a factor of x10. The significance of our approach is in fortifying the security posture of IoT devices and maintaining a robust implementation of trustful healthcare systems.
\end{abstract}


\section{Introduction}
\begin{figure}[htbp]
\centerline{\includegraphics[width =\linewidth]{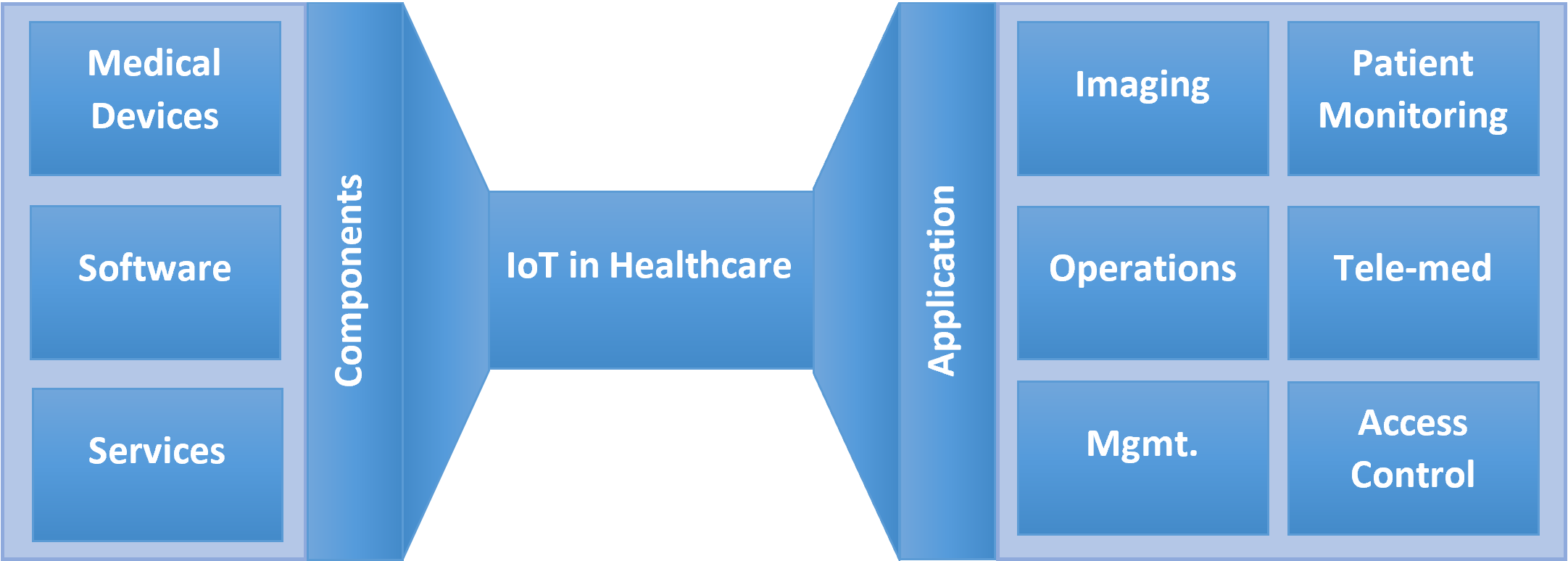}}
\caption{IoT Components Share in Healthcare.}
\label{fig1}
\end{figure}

\begin{figure}[htbp]
\centerline{\includegraphics[width =\linewidth]{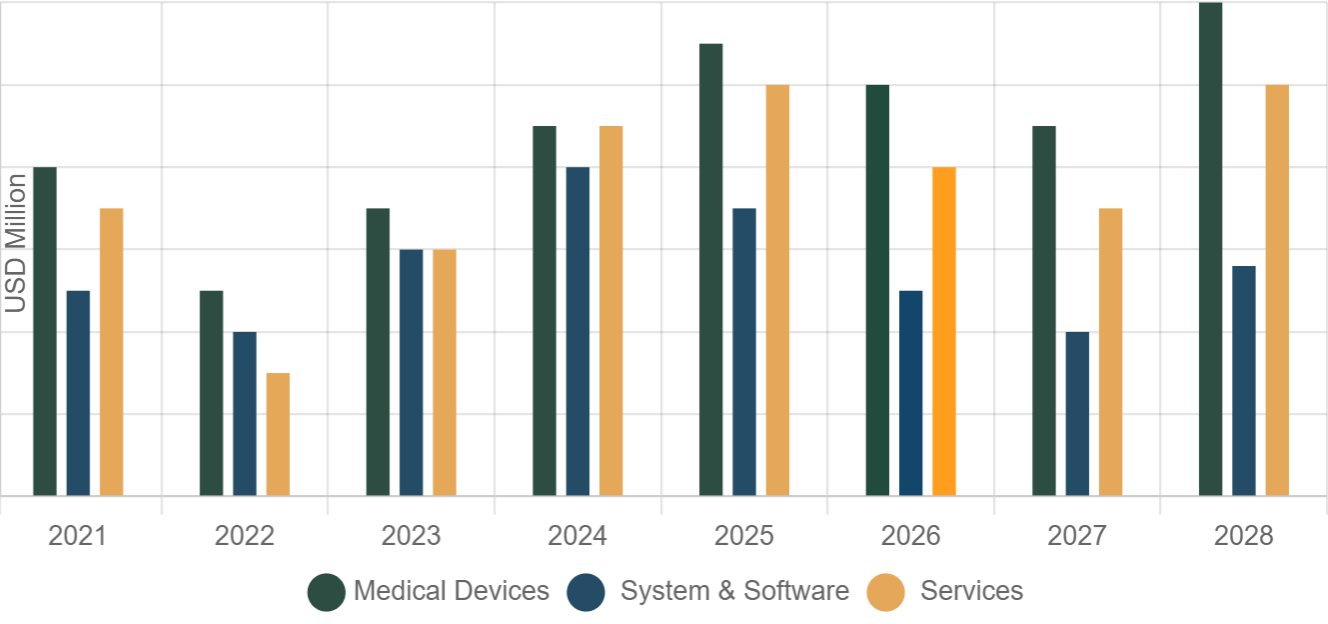}}
\caption{IoT in Healthcare Market, by Component (USD M), 2019-2028 [1].}
\label{fig2}
\end{figure}

Integrating the Internet of Things (IoT) devices in healthcare applications has ushered in a new era of patient-centric care, real-time monitoring, and data-driven decision-making. The Global IoT in Healthcare Market value is estimated at USD 252.2 Billion in 2023~\cite{cybsec}. The size of this market is expected to grow at a rate of CAGR 16.9\% to USD 550.58 Billion in 2028. The Global IoT in Healthcare Market is segmented by component into Medical Devices, Systems and Software, and Services can be seen in Fig.~\ref{fig1}, and the predicted market share is shown in Fig.~\ref{fig2}.

This paradigm shift, however, brings with it a set of unprecedented challenges, particularly in the domain of security. The healthcare sector, characterized by its sensitive and confidential nature, is increasingly dependent on IoT devices for tasks ranging from remote patient monitoring to managing medical equipment and facilitating data-driven diagnoses as well as device availability.

As the healthcare industry undergoes a digital transformation with the widespread adoption of Internet of Things (IoT) devices, the urgency to implement proactive security measures becomes increasingly critical. The predicted cost of cybercrime to the world economy in 2020 was less than \$1 trillion, a rise of more than 50\% from 2018~\cite{Mahdavifar2020}. The interconnected nature of IoT devices introduces a complex attack surface, leaving healthcare systems susceptible to a myriad of potential threats. The landscape of cyber threats is constantly evolving, with threat actors becoming more sophisticated and resourceful. In recent years, the healthcare sector has witnessed an alarming surge in cyber-attacks, ranging from distributed denial of service (DDOS), Recon, Greeth flood, UDPPlain, and ransomware targeting medical records to the compromise of connected medical devices. These attacks can jeopardize patient data privacy as well as it can potentially disrupt healthcare services availability, putting patient lives at risk.

Traditional security approaches, often based on predefined signatures and rule-based systems, need help to keep pace with the unpredictability of emerging threats. Cyber adversaries frequently exploit unknown vulnerabilities, necessitating a shift from reactive to proactive security measures, in the healthcare IoT ecosystem, where the stakes are particularly high due to the sensitive nature of patient information and the criticality of medical devices, the traditional security paradigm needs to provide comprehensive protection. The urgency to address security concerns is further heightened by the fact that healthcare systems often cannot afford extended periods of downtime or disruption.

Proactive security measures involve a shift from reactive incident response to anticipatory risk mitigation. By leveraging advanced technologies such as machine learning, organizations can identify and address potential vulnerabilities before they are exploited. This proactive approach is crucial in the healthcare IoT context, where the timely identification and mitigation of risks can prevent data breaches, ensure the continuous operation of medical devices, and safeguard patient well-being.
This paper presents a comprehensive examination of the security landscape within healthcare IoT, identifies existing challenges, and introduces an innovative ML-based architecture designed to address these challenges while achieving up to 93.6\%  accuracy and reducing the cost of the monitor sensors by a factor of x10 to fortifying the security requirement of IoT devices in healthcare applications.

\section{Background}

Internet of Things (IoT)-enabled devices have made remote monitoring in the healthcare sector possible, unleashing the potential to keep patients safe and healthy and empowering physicians to deliver excellent care. As communication between patients and physicians has gotten simpler and more effective, it has also raised patient happiness and engagement. In addition, remote patient monitoring minimizes hospital stays and keeps patients from being readmitted. IoT has a significant influence on both increasing treatment outcomes and drastically lowering healthcare expenditures.

\subsection{IoT in Healthcare}

\begin{figure}[htbp]
\centerline{\includegraphics[width =\linewidth]{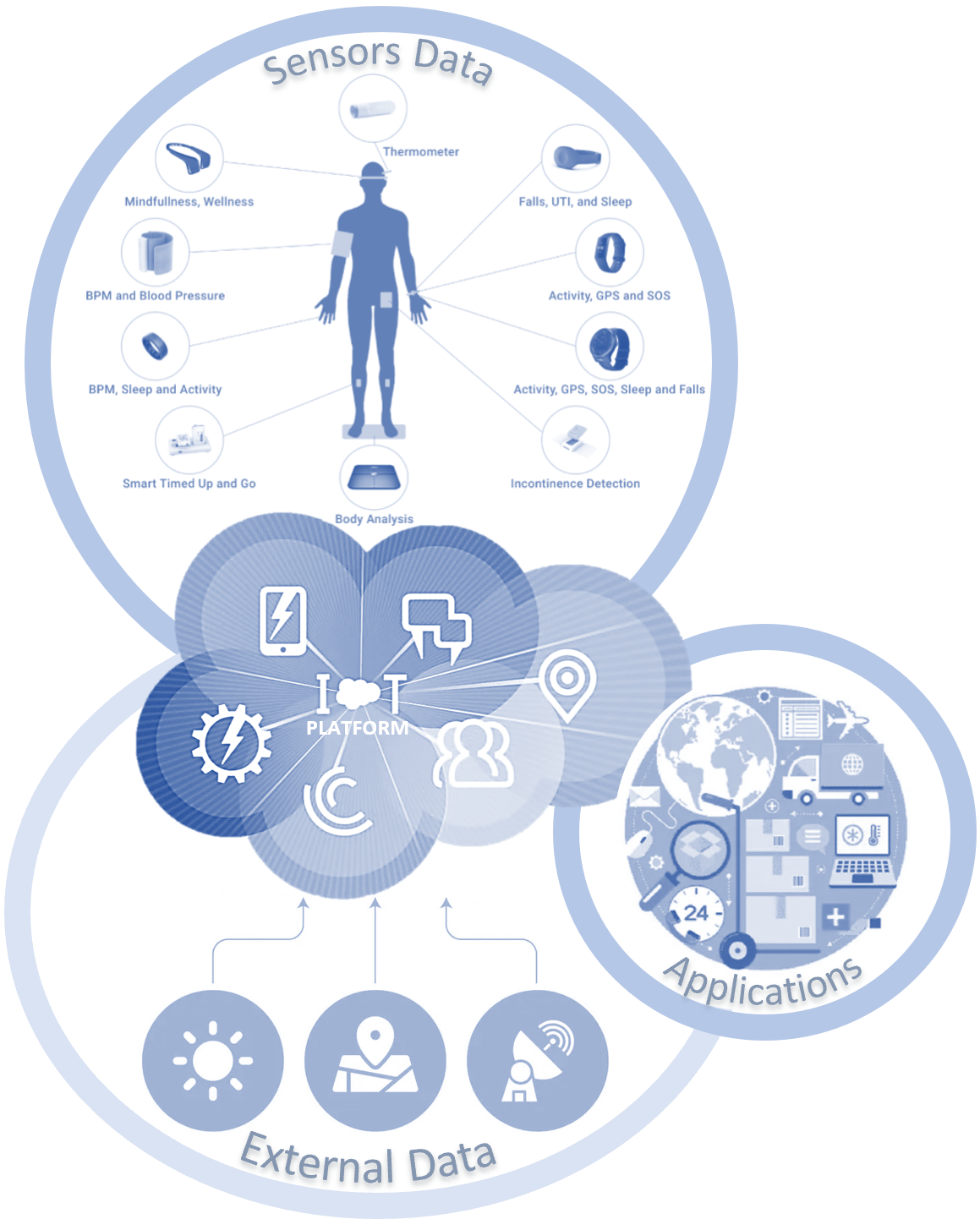}}
\caption{Healthcare IoT framework block diagram.}
\label{fig3}
\end{figure}

The usage of the Internet of Things (IoT) in healthcare has ushered in a transformative era, enhancing patient care, optimizing operational efficiency, and revolutionizing how healthcare providers deliver services, such as and not limited to Remote Patient Monitoring, Connected Medical Devices, Healthcare Asset Tracking, Smart Hospitals and Infrastructure, Medication Management, Smart Pills, Telehealth and Telemedicine, Patient Engagement, and  Data Analytics~\cite{kashani2021systematic}. An abstract healthcare IoT framework block diagram is shown in Fig.~\ref{fig3}.

\subsection{IoT Framework Anatomy}
\textit{Communication}: The Internet of Things (IoT) allows heterogeneous items to communicate with one another via a variety of communication protocols, including Bluetooth, IEEE 802.15.4, NFC, Wi-Fi, RFID, ultra-wide bandwidth (UWB), Z-wave, and long-term evolution (LTE), etc.

\textit{Computation}: The computing power of the Internet of Things is demonstrated via processing units (microprocessor, microcontroller, field-programmable gateway array, system on a chip), hardware platforms (Raspberry Pi, Mulle, T-Mote Sky, Arduino, Intel Galilio, UDOO, etc.), and software platforms. The operating system operates while the system is being activated, is the most significant software platform. The creation of Internet of Things applications frequently makes use of real-time operating systems (RTOS)~\cite{stankovic}. 

\textit{Services}: Identity-related, omnipresent, collaboratively aware, and information aggregation services are the different categories under which IoT services fall. The fundamental services that are essential to other services' functionality are identity-related services~\cite{ghosh2010lte}. Applications must detect real-world items in order to display them in the virtual environment. The sensor data is gathered, processed, and reported to IoT applications by the information aggregation services. Services that are aware of collaboration make decisions and respond appropriately based on data from services associated with aggregate.

\textit{Metadata}: This is the ability to extract information from different devices in the IoT and then offer the required services to the user. Extraction contains discovering information, uses of information, and modeling information.

\subsection{Significance of Cybersecurity in Healthcare}
In~\cite{alasmari2016security}, the work was specially created with security and privacy in mind for the Internet of Things. The writers talk about how the Internet of Things' cloud created a dangerous issue for the healthcare industry. In this study, recommendations were provided to encourage the researcher to address the concerns regarding the security and privacy of healthcare systems in the context of the Internet of Things.

As discussed in the~\cite{SCHILLER2022100467}, the limitations of traditional security measures, often characterized by static rule-based systems, face significant limitations in the context of healthcare IoT. The deterministic nature of rule-based systems proves inadequate in addressing the dynamic and evolving nature of cyber threats. Moreover, these approaches struggle to adapt to the heterogeneity of IoT devices, each with its unique communication protocols, data formats, and security requirements. This reactive approach, while effective for known vulnerabilities, falls short in addressing zero-day attacks and emerging threats unique to healthcare IoT.

Recognizing the limitations of traditional security measures and the evolving threat landscape, there is an imperative to transition from reactive to proactive security measures in healthcare IoT. Proactive security involves anticipatory risk mitigation, continuous monitoring, and adaptive responses based on real-time analysis of data patterns.

\subsection{IoT in Healthcare Challenges }
In~\cite{darshan2015}, the author discusses the prediction of early symptoms through the IoT, which is useful for the patient and a very beneficial tool for society. It discusses all the papers related to reviews and challenges associated with the IoT in the healthcare system, which can be summarized to:
\begin{itemize}
    \item Heterogeneity of Devices.
    \item Scale and Volume.
    \item Resource Constraints.
    \item Distributed Nature.
    \item Dynamic Nature.
    \item Data Privacy Concerns.
    \item Standards.
    \item Long Life-cycle.
    \item Edge and Fog Computing.
\end{itemize}

The complete comparison of challenges in traditional and Health IoT security systems is shown in Table.~\ref{tab1a}.

\subsection{Machine Learning Approach}
Machine Learning emerges as a catalyst in the paradigm shift towards proactive security in healthcare IoT~\cite{WAZID2022313}. ML's capacity to analyze vast datasets, identify patterns, and autonomously adapt to emerging threats seamlessly with the dynamic nature of healthcare IoT environments. ML algorithms include but are not limited to anomaly detection, predictive analytics, dynamic threat intelligence, adaptive access control, incident response automation, and behavioral analysis.

Additionally, healthcare IoT systems often deal with diverse data and user profiles~\cite{math11122758}. ML enables the customization of security measures based on individualized risk assessments. By understanding user behavior and access patterns, ML models can dynamically ensure that the right level of protection is applied to different devices and users within the healthcare ecosystem. 

\begin{table}[htbp]
\caption{Security Concerns Compassion between Traditional and IoT Systems.}
\begin{center}
\begin{tabular}{|p{0.12\linewidth}|p{0.35\linewidth}|p{0.35\linewidth}|}
\hline
\textbf{Concern}&\multicolumn{2}{|c|}{\textbf{Security Concerns}} \\
\cline{2-3} 
\textbf{Point} & \textbf{\textit{Traditional}}& \textbf{\textit{IoT}} \\
\hline
Hetero-
geneity&Designed for homogeneous environments with standardized systems. One-size-fits-all approach.&IoT ecosystems consist of diverse devices with varying capabilities, communication protocols, and security postures. Integrating and securing a myriad of devices with different architectures poses a considerable challenge.  \\
\hline
Scale and Volume&Struggle to scale efficiently to handle the sheer magnitude of data.& Generate vast amounts of data, challenging to manage and analyze in real-time. Scaling security measures to accommodate the growing number of connected devices.  \\
\hline
Resource Constraints&Assumes the availability of ample resources, which may not be the case for resource-constrained IoT devices. &Limited processing power, memory, and energy. Implementing robust security measures within these constraints requires specialized approaches.  \\
\hline
Distributed Nature&More centralized and focused on securing specific endpoints within a local network. &Inherently distributed, with devices often communicating across networks and ecosystems. Securing the entire communication chain and ensuring end-to-end security is a complex undertaking.  \\
\hline
Dynamic Environments& often static and may struggle to adapt quickly to the dynamic nature of IoT ecosystems. &Dynamic, with devices joining and leaving the network frequently. This dynamism requires security measures that can adapt to changes in the device landscape.  \\
\hline
Data Privacy &While traditional security models address data privacy concerns, the scale and nature of data in IoT systems introduce additional complexities. & IoT devices often collect and transmit sensitive data, including personal and health-related information. Safeguarding this data against unauthorized access and ensuring privacy is a critical challenge.  \\
\hline
Standards&Traditional security frameworks often rely on established standards, making it easier to implement consistent security measures. &The absence of standardized security protocols across all IoT devices creates interoperability challenges and may result in vulnerabilities if not properly addressed.\\
\hline
Long Lifecycle&Systems may not have been designed with the long lifecycles typical of many IoT devices, leading to potential obsolescence. &Many IoT devices have extended lifecycles, and security measures must be designed to withstand evolving threats over an extended period.\\
\hline
Edge Fog Computing&Traditional security models may not have originally considered the decentralized nature of edge and fog computing. &With the rise of edge and fog computing in IoT architectures, security measures must extend beyond centralized servers to include distributed computing nodes, adding complexity to the security .\\
\hline
\end{tabular}
\label{tab1a}
\end{center}
\end{table}

Furthermore, Zero-day attacks based on exploiting vulnerabilities unknown to security experts pose a significant threat in the healthcare IoT domain~\cite{GUO2023175}. ML's ability to recognize patterns without relying on predefined signatures allows for rapid identification of potential zero-day threats. This enables swift responses and the implementation of countermeasures to protect against emerging and previously unseen security risks. Detailed research has been conducted on using the ML model for zero-day attack detection~\cite{HUDA2017211}~\cite{electronics9101684}~\cite{Mirsky2018KitsuneAE}. Various ML models, ranging from unsupervised ML and supervised ML to Transfer Learning, are explored. In the study in~\cite{WAZID2022313}, the authors provide a detailed discussion about the different concepts by uniting cybersecurity and ML. 

\subsection{Related Work}
Several IoT systems were proposed for healthcare facilities in hospitals. In~\cite{usability}, a real-time monitoring intensive care unit (ICU) is proposed. The authors gave an interactive environment with suggestions for hardware and services offered by the system. The system has yet to be made available for clinical usage; it is currently in the testing and validation stages. Additionally, there is still work to be done to secure the communication channels and threat protection.

In~\cite{overview} proposed a lab-created system for asthma patients that suggests a method of measuring patients' vital signs (such as their temperature, respiration rate, and number of breaths) that uses signal augmentation and watermarking techniques to ensure secure data transmission. However, there are several shortcomings in using commercial wearable IoT devices and their inability to protect from DDoS attacks. 

In~\cite{s23156760}, the authors suggest a brilliant description of a cloud-based patient monitoring approach that gathers, stores, processes, and visualizes IoT data that comprise the three primary components of system security, where the vast amount of IoT data, the sensing module uses variable sampling and filtering. However, the architecture depends only on isolating the traffic via separate gateways to secure the system.

\begin{figure}[htbp]
\centerline{\includegraphics[width =\linewidth]{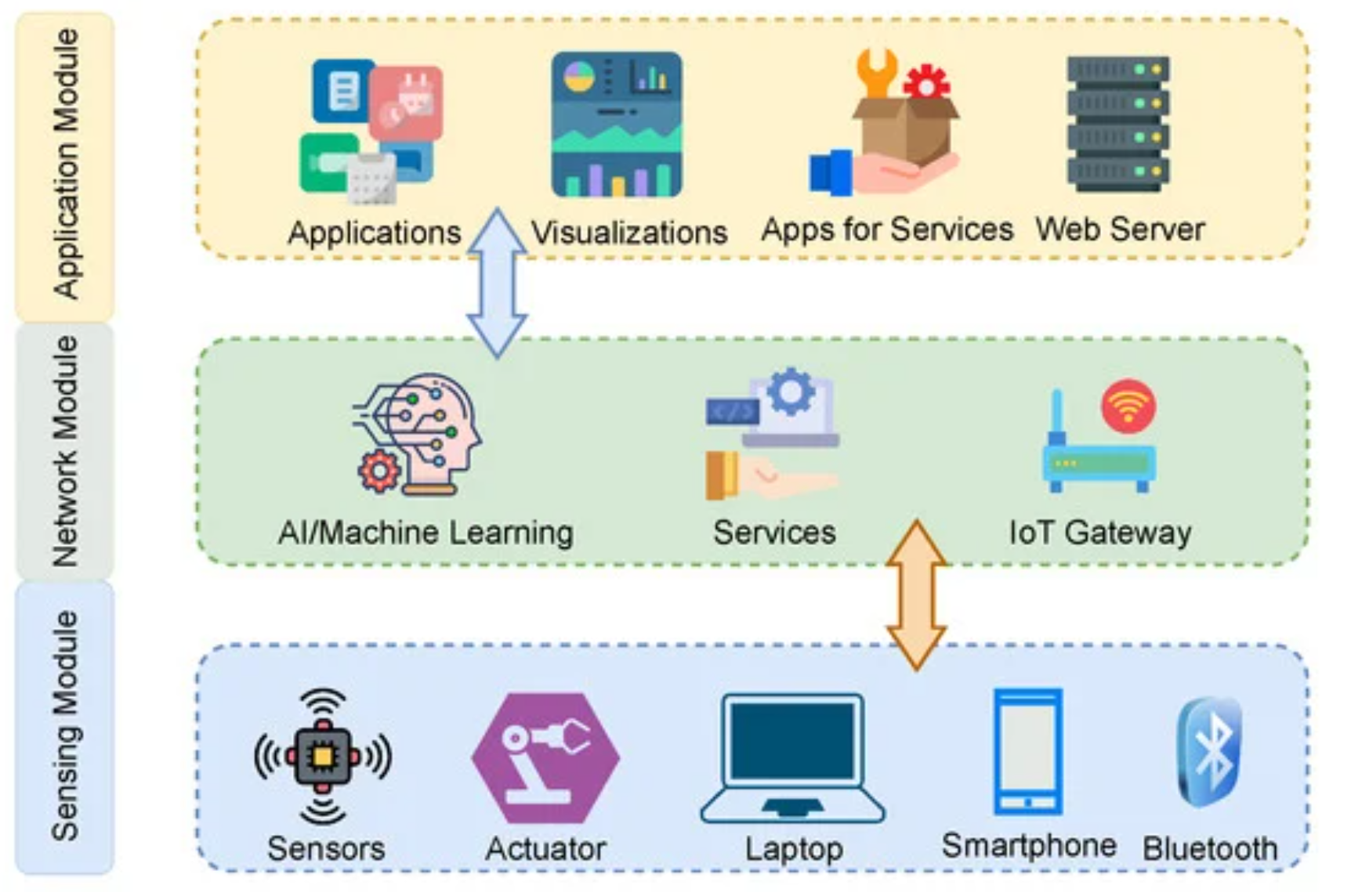}}
\caption{A block diagram of the common IoT Healthcare architecture model~\cite{nasiri}.}
\label{fig4}
\end{figure}

In~\cite{nasiri}, a tremendous effort was made to survey the most remarkable works on IoT implementation and their security focus while highlighting and classifying the requirements into confidentiality, integrity, availability, authentication, authorization, accountability, data freshness, non-repudiation; and discussing the resiliency requirement of reliability, Repair-ability, adaptability, safety, and performance. 
Almost all the related work focuses on one or more concerns of the IoT cybersecurity for health care and follows the framework, shown in Fig.~\ref{fig4}.

In order to significantly improve security and performance for wireless body area networks (WBANs) in~\cite{NYANGARESI2023103117}, the authors provide a revolutionary three-factor authentication protocol that integrates patient bio-metrics, smart cards, and passwords was created. Additionally, The system results in a 43.98\% reduction in computation overheads; however, it increases the cost of the implementation dramatically.

\section{Autoencoder}
\begin{figure}[htbp]
\centerline{\includegraphics[width =\linewidth]{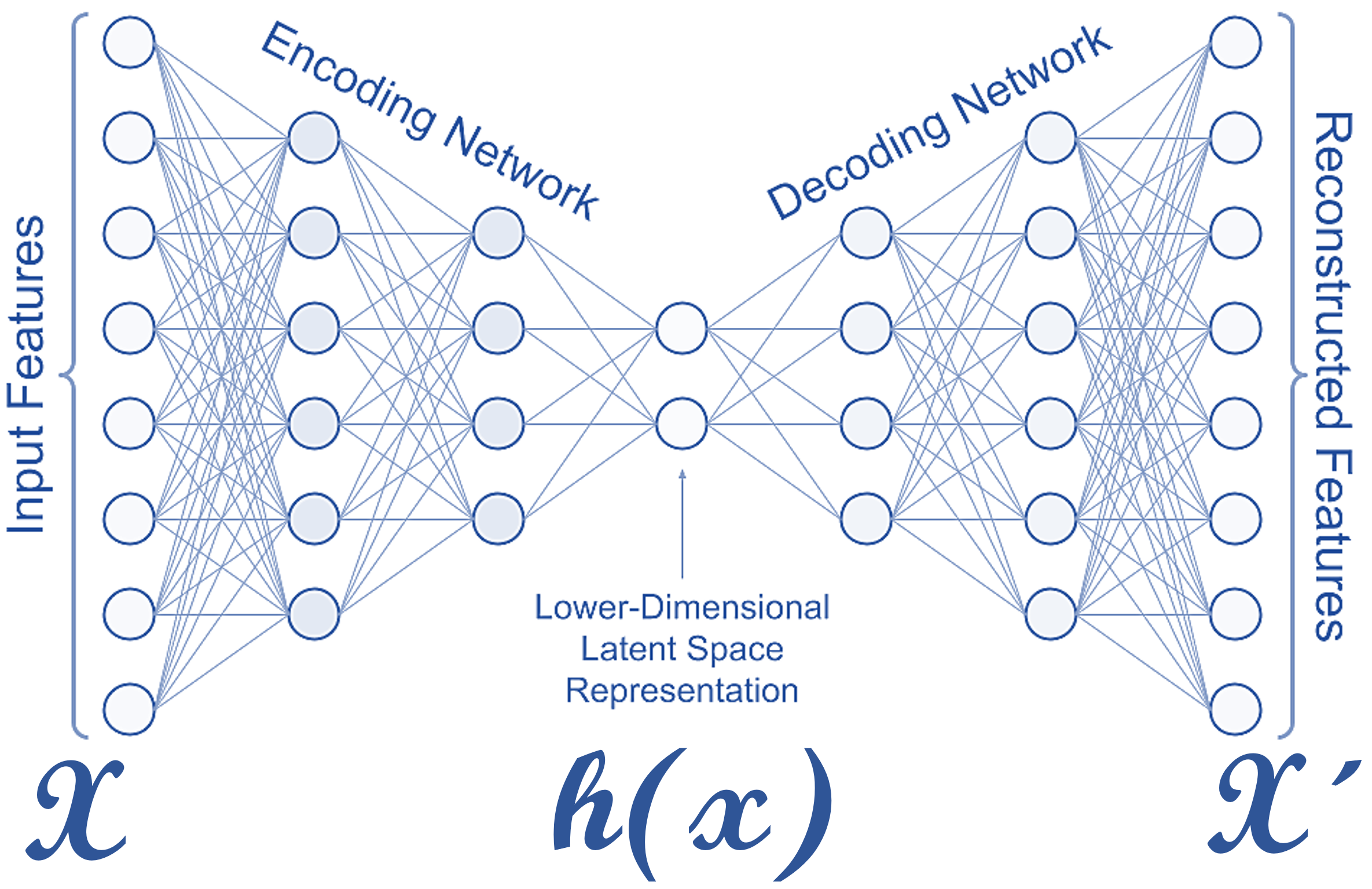}}
\caption{Network Architecture for an abstract Autoencoder ML model.}
\label{fig5}
\end{figure}

An artificial neural network called an autoencoder is utilized for unsupervised learning or the efficient coding of unlabeled input. An autoencoder~\cite{LI2023110176} picks up two new skills: decoding, which reconstructs the input data from the encoded representation, and encoding, which modifies the input data that is very beneficial for zero-day attacks; the autoencoder structure is shown in Fig.~\ref{fig5}.

Depending on the domain and use case, an autoencoder design and the number of hidden layers vary, and the principal model formula is listed below.

\begin{equation}
\Phi+\Psi= \arg \min_{\Phi \Psi} || X-(\Phi \odot \Psi)X'||^2\label{eq1}
\end{equation}
\noindent
where $x$ is the input, $x'$ is the output, $Phi$ is the encoding function $Psi$ is the decoding function. The function that represents the variation between the input $x$ and the reconstructed input $x'$ defines the reconstruction error. \eqref{eq2} and \eqref{eq3} illustrate the usage of mean square error (MSE) and mean absolute error (MAE), two popular functions, to determine the reconstruction error for a sample size of $N$.
\begin{equation}
MSE= \sum_{i=1}^{N} (x'-x)^2 \\ \label{eq2}
\end{equation}

\begin{equation}
MAE= \sum_{i=1}^{N} |x'-x| \\ \label{eq3}
\end{equation}

As it is noticed from the previous work and to the best of our knowledge, there is no dedicated network security monitoring (NSM) with intrusion prediction capabilities that was implemented specifically targeting IoT system security. Keeping in mind the fact that all security systems will eventually fail, they were not monitored accordingly. Thus, in this work, we propose a novel architecture tailored to IoT needs and a robust implementation of IoT ML-based NSM devices.

\section{Proposed Architecture}
The proposed system must inject the NSM concept into the traditional IoT framework, providing a secure IoT architect for Healthcare applications. In addition to crafting the novel component into a simple add-on hardware device that could be added to any exciting IoT healthcare implementation. 

The proposed architecture consists of two main components: the sensor monitor (SM) and the intrusion detection hub (IDH). The SM unit consists of a miniature network sensing device specialized in IoT traffic only by filtering all inbound and outbound connections over the WiFi Network and Zigbee signaling, limiting the captured traffic to the protocols listed in Table~\ref{tab1e2}.

\begin{table}[htbp]
\caption{The filtered protocols for the Signal Monitoring Unit.}
\begin{center}
\begin{tabular}{|c|c|c|}
\hline
 \textbf{\textit{Protocol}}& \textbf{\textit{Description}}& \textbf{\textit{TCP or UDP}} \\
\hline
AMQP&Advanced Message Queuing Protocol&TCP(5671, 5672)\\
\hline
CoAP&Constrained Application Protocol&UDP(5683) \\
\hline
LWM2M&Lightweight M2M&UDP (5783,5784)  \\
\hline
MQTT&Lightweight publish-subscribe M2M&TCP (1883, 8883)  \\
\hline
XMPP&Extensible Messaging Presence Protocol&TCP (5269, 5280)  \\
\hline
\end{tabular}
\label{tab1e2}
\end{center}
\end{table}

\begin{figure}[htbp]
\centerline{\includegraphics[width =\linewidth]{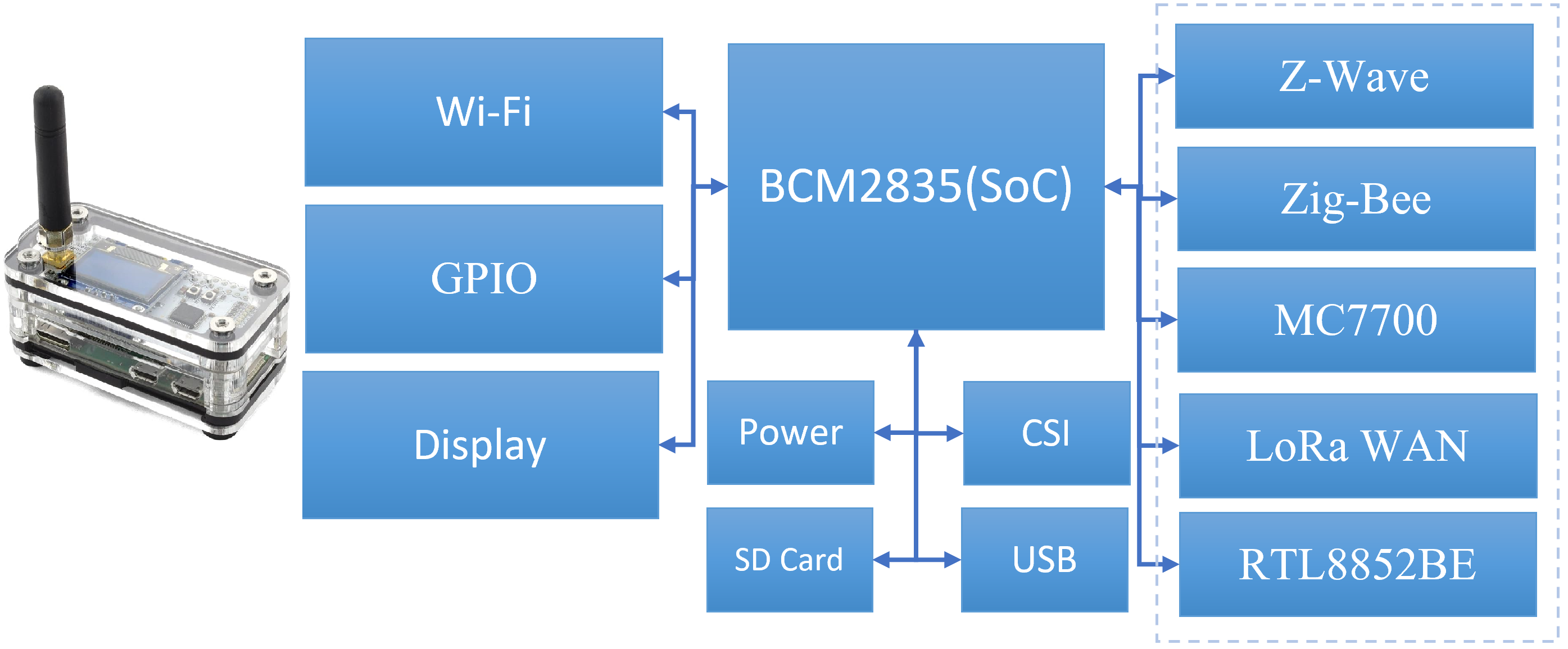}}
\caption{The SM unit hardware block diagram and the visual ZRZ-1S packing.}
\label{fig6}
\end{figure}

The SM and the IDH units are equipped with several antennas to accommodate the most popular physical layer protocols that will be used in healthcare IoT devices by 2023; the currently supported protocol and the associated modules are listed in Table~\ref{tab1e3}. Thanks to the modular design concept used in our proposed architecture, the units can be simply modified, adjusted and upgraded to any future signal technology. The block diagram of the SM unite is demonstrated in Fig.~\ref{fig6}. 

The SM unit acts as an agent proxy to collect IoT filtered traffic to the main security monitoring hub IDH, which consists of a SoC board with a preloaded ML prediction module that was trained for various IoT threats, inspired by the algorithm presented in~\cite{electronics9101684} autoencoder algorithm is used where an An Artificial Neural Network (ANN) serves is the foundation, and the network design, epoch count, and learning rate are chosen for hyper-parameter optimization via the technique described~\cite{bergstra2012random} and the number of layers and node was reduced by 20\% to accommodate the hardware implementation limits. It is well known that random search converges to a semi-optimal set of parameters more quickly than grid search. When few criteria are required, it has also been demonstrated to perform better than grid search and reduces the likelihood of getting parameters that are over-fitted. For training, testing, and verification, Initially, we used the dataset CICIoT2023's data from seven categories, namely DDoS, DoS, Recon, Web-based, Brute Force, Spoofing, and Mirai. The input parameters are flow duration, Header Length, Protocol Type, Duration, Rate, Srate, Drate, flagnumber, ack count, syn count, fin count, urg ount, rst count, HTTP, HTTPS, DNS, Telnet, SMTP, SSH, IRC, TCP, UDP, DHCP, ARP, ICMP, IPv, LLC, and Tot sum. The input dataset was filtered for the outliers, and then the .pcap files were processed to generate bidirectional traffic and then saved into .csv file format. Finally, the dataset was divided into 75\% training and 25\%.

\begin{table}[htbp]
\caption{The filtered physical layer antennas protocols for SM.}
\begin{center}
\begin{tabular}{|c|c|}
\hline
 \textbf{\textit{Protocol}}& \textbf{\textit{Hardware}}  \\
\hline
Bluetooth&Realtek RTL8852BE-CG \\
\hline
Cellular&Sierra Wireless MC7700  \\
\hline
LoRaWAN&SX1262 LoRa HAT 915MHz  \\
\hline
Wi-Fi& Built in BCM4345  \\
\hline
Zigbee&XB24CAWIT \\
\hline
Z-Wave&ZMEURAZ2 \\
\hline
\end{tabular}
\label{tab1e3}
\end{center}
\end{table}

\begin{figure}[htbp]
\centerline{\includegraphics[width =\linewidth]{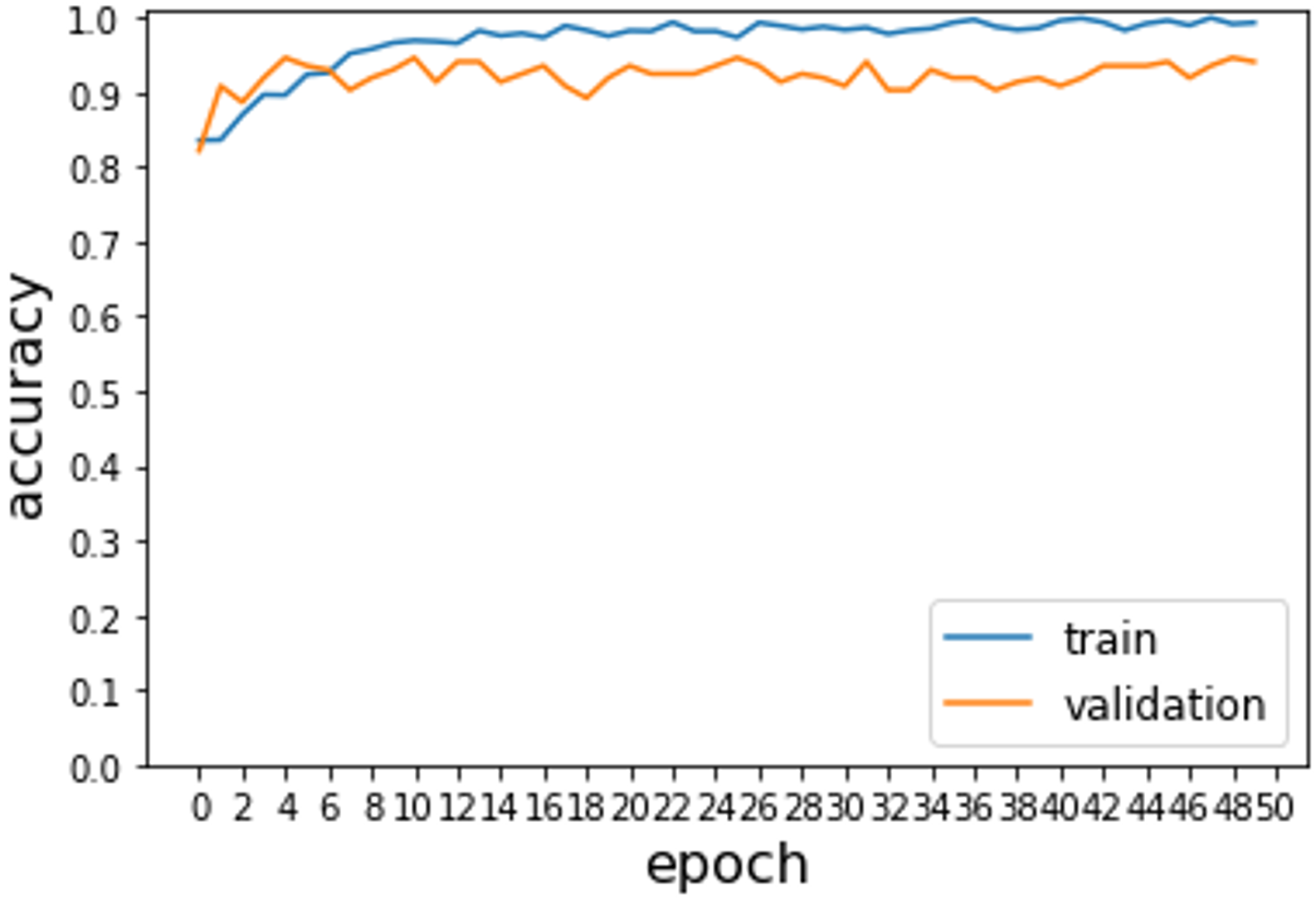}}
\caption{The IDH unit core prediction module training and validation graph.}
\label{fig7}
\end{figure}

The model was implemented in Python relying on TensorFlow 2.10.1 and Keras 2.10 libraries, and it works as follows: If the Mean Squared Error (MSE) between the original and decoded instances is more than a certain threshold, the attack instance is considered a zero-day attack. Based on heuristic try and error, the system was tested for the range [0.05: 0.95] as the parameter optimization; the value 0.15 was chosen to converge the MSE; it is a crucial factor in determining the value at which an instance is classified as a zero-day assault, where the convergence graph of 50 epochs is shown in Fig.~\ref{fig7}. 

The IDH unit is implemented using a Raspberry Pi 4B to reduce the cost and power consumption. Where the pick power consumption reached 7.5W at an accuracy up to 93.6\%, producing 4.7 mgCO\textsubscript{2} comparing to at least 725.9 mgCO\textsubscript{2} that is produced by the minimal equivalent server gives the current technology.

\section{Comparison and Discussion}
We compared the proposed architecture to the most coupled systems such as(Sys1)~\cite{s23156760}, (Sys2) ~\cite{peppes2023}, and (Sys3)~\cite{Gharib2019} and as the proposed method is novel and there was no exact analogy as of our best knowledge. The accuracy, power consumption, dataset, and hardware implementation were used as comparison parameters. As shown in Table~\ref{tab1e4}, the proposed method's accuracy exceeds the state-of-the-art hardware implementation while competing with the software developed that uses the same model approach but benefits from extra hardware power.

\begin{table}[htbp]
\caption{The result comparison to (Sys1)~\cite{s23156760}, (Sys2) ~\cite{peppes2023}, and (Sys3)~\cite{Gharib2019}.}
\begin{center}
\begin{tabular}{|c|c|c|c|c|}
\hline
 \textbf{\textit{Param}}& \textbf{\textit{Sys1}} & \textbf{\textit{Sys2}}  & \textbf{\textit{Sys3}} & \textbf{\textit{Proposed}}\\
\hline
Accuracy (\%)&98.4&96.4&90.17&\textbf{93.6} \\
\hline
Algorithm &Autoencoder&GANs&AutoIDS&\textbf{Autoencoder} \\
\hline
Power (W)&--&--&N.A.&\textbf{7.5}  \\
\hline
DataSet&CICIDS2017&Kaggle&NSL-KDD&\textbf{CICIOT2023}  \\
\hline
Hardware&--&--&RPi&RPi4  \\
\hline
\end{tabular}
\label{tab1e4}
\end{center}
\end{table}

The authors are aware that the system is performing with 93.6\% mean accuracy with 82.1\% worst-case scenario according to the simulation, and we believe that with a more extensive dataset, the standard deviation of the accuracy can be further improved. Additionally, the power consumption is relatively low; however, the module puts the SoC to operate at 98\% utilization, which can stress the hardware board if it will be operating in high ambient temperature; however, in a healthcare setup, this scenario is very unlikely.

\section{Conclusion}

This study has thoroughly investigated the difficulties, requirements, and creative solutions related to healthcare Internet of Things (IoT) system security. The combination of IoT and healthcare technology has the potential to revolutionize patient care. Still, it also requires the use of cutting-edge solutions and a thorough grasp of the complex cybersecurity landscape. The proposed (ML)-based architecture strengthens the security posture of Internet of Things (IoT) systems in the healthcare industry. This architecture encompasses a range of features, such as adaptive threat detection, real-time anomaly detection, and predictive risk mitigation, while competing with the full software-developed solutions and leading the hardware-implemented stat-of-the-art with the accuracy of 93.6\% with a low-cost solution that generates producing 4.7 mgCO\textsubscript{2} only making the proposed solution and environment-friendly architecture.

\bibliographystyle{ieeetr}
\bibliography{references}

\end{document}